# An Effective Automatic Image Annotation Model Via Attention Model and Data Equilibrium

Amir Vatani*
Department of Computer Engineering, Islamic Azad University, Science & Research Branch of Tehran, P.O. Box 14515/775, Iran

Milad Taleby Ahvanooey
School of Computer Science and Engineering, Nanjing University of Science and Technology, Nanjing, P.O. Box 210094 P.R. China

Mostafa Rahimi
Department of Electrical and Computer Engineering,
Shahid Beheshti University G.C, Tehran, Iran

*Abstract*—Nowadays, a huge number of images are available. However, retrieving a required image for an ordinary user is a challenging task in computer vision systems. During the past two decades, many types of research have been introduced to improve the performance of the automatic annotation of images, which are traditionally focused on content-based image retrieval. Although, recent research demonstrates that there is a semantic gap between content-based image retrieval and image semantics understandable by humans. As a result, existing research in this area has caused to bridge the semantic gap between low-level image features and high-level semantics. The conventional method of bridging the semantic gap is through the automatic image annotation (AIA) that extracts semantic features using machine learning techniques. In this paper, we propose a novel AIA model based on the deep learning feature extraction method. The proposed model has three phases, including a feature extractor, a tag generator, and an image annotator. First, the proposed model extracts automatically the high and low-level features based on dual tree continues wavelet transform (DT-CWT), singular value decomposition, distribution of color ton, and the deep neural network. Moreover, the tag generator balances the dictionary of the annotated keywords by a new log-entropy auto-encoder (LEAE) and then describes these keywords by word embedding. Finally, the annotator works based on the long-short-term memory (LSTM) network in order to obtain the importance degree of specific features of the image. The experiments conducted on two benchmark datasets confirm that the superiority of proposed model compared to the previous models in terms of performance criteria.

## I. INTRODUCTION

Automatic image annotation (AIA) is one of the image retrieval techniques in that the images can be retrieved in the same way as text documents. In the AIA, the main idea is to automatically learn the semantic concept models from a huge number of image samples and utilize the conceptual models to label new images with proper tags [1]. The AIA has a lot of applications in various fields including access, search, and navigate the huge amount of visual data which stored in online or offline data sources, image manipulation and annotation application that used on a mobile device, [2]-[4]. The typical image annotation approaches rely on human viewpoints and the performance of them is highly dependent on the inefficient manual operations. Recently, many types of research [5], [6], [12], [31] have been conducted on the AIA that can be grouped into two different models [13]; generative models, such as [1], [7], and discrimination or conditional models such as [3], [12]. The generative models try to learn the joint probability distribution between keywords and image features [8], [13]. Simultaneously, conditional models are a class of models used in machine learning for modeling the dependence of semantic keywords on visual features [8]. During the last decade, deep learning techniques have reached excellent performance in the field of image processing. Furthermore, visual attention with deep neural networks has been utilized successfully in many natural languages processing and computer vision systems. It also has been used for image annotation issue in some existing literature [17], [19], [24], [34]. Although the existing deep learning based techniques have improved the performance of AIA models, still there are two major limitations including management of imbalanced distribution keywords and selection of correct features. To address these problems, we propose a technique for extracting the high-level and low-level features that are able to extract them automatically based on dual tree continues wavelet transform (DT-CWT), singular value decomposition, distribution of color ton and the deep neural network. Next, we utilized an attention model for weighting the important feature by considering suitable coefficient. Moreover, we suggested a tag generator that works based on the log-entropy auto-encoder, and LSTM networks and then treat each keyword equally, in imbalanced distribution dictionary in order to find the better similar tags (e.g., keywords are described by word embedding approach).

The rest of this paper is organized as follows. Section II presents a brief description of existing literature on image annotation. Section III presents the proposed AIA model. Section IV discusses the experimental results and compares the proposed AIA model with the state of art techniques. Section V draws some conclusions.

## II. Related Work

In this section, we introduce some existing literature on AIA. During the last two decades, the AIA has been an active research area in the field of pattern recognition and computer vision. several AIA techniques have been proposed to improve the performance of AIA models, which some of them try to learn the joint probability distribution between keywords and image features called generative models [1], [3]. In addition, some other techniques treat based on supervised learning problem in order to overcome the issue of image annotation, which is named discrimination models [1], [3]. Furthermore, some existing techniques have utilized a combination of these two methods; for example, in visual object classification, the combination of generative and discrimination model has been used. However, the difference between the AIA and the classification task is that each sample always has multiple correlated annotations, which makes it difficult to apply the combination of generative and discrimination techniques for the AIA [13].

Ping et al. [1] combined the generative and discriminative models by local discriminant topics in the neighborhood of the unannotated image by applying the singular value decomposition grouped the images of the neighborhood into different topics according to their semantic tags.

Mei Wang et al. [3] suggested an AIA model via integrated discriminative and generative models. This model first identifies a visual neighborhood in the training set based on generative technique, and then, the neighborhood is defined by an optimal discriminative hyperplane tree classifier based on the feature concept. The tree classifier is generated according to a local topic hierarchy, which is adaptively created by extracting the semantic contextual correlations of the corresponding visual neighborhood.

Duygulu et al. [33] expressed an object recognition model as machine translation. In that model, the recognition is a process of annotating image regions by words. In addition, it utilizes a translation model to form the relations between the image visual words and words in order to label new images. This model provides a possibility to extract features by some approaches, which tries to select proper features for improving the performance of recognition.

Dongping et al. [5] proposed a new AIA model based on Gaussian mixture model (GMM) by considering cross-modal correlations. In this model, first, the GMM is fitted by the rival penalized competitive learning (RPCL) or expectation-maximization algorithm in order to predict the posterior probabilities of each annotation keyword. Moreover, an annotation similarity graph is generated with a weighted linear combination of visual similarity and label similarity by integrating the information from both high-level semantic concepts and image low-level features together. The most important merit of this model is that it is able to effectively avoid the phenomena of synonym and polysemy appeared during the annotating process.

Song et al. [12] introduced a Sparse Multi-Modal Coding for image annotation using an efficient mapping method, which functions based on stacked auto-encoders. In this work, they utilized a new learning objective function, which obtains both intra-modal and semantic relationships of data from heterogeneous sources effectively. Their experimental results conducted on some benchmark datasets demonstrate that it outperforms the baseline models for the task of image annotation and retrieval.

Wang et al. [27] provided a new image annotation method by focusing on deep convolutional neural network for large-scale image annotation. They contacted the proposed method on the MIRFlickr25K and NUS-WIDE datasets in order to analyze its performance. In practice, this method analyzes a pre-specify dual-model learning scheme which consists of learning to fine-tune the parameters of the deep neural network with respect to each individual modality and learning to find the optimal combination of diverse modalities simultaneously in a coherent process.

Karpathy and Fei-Fei [35] presented a novel AIA model that produces natural language descriptions of images and their regions based on the weak labels by performing on a dataset of images and sentences (e.g., with respect to very few hardcoded assumptions). This model employs the leverages images and their sentence descriptions in order to learn about the inter-modal correspondences between language and visual features. In the results, they evaluated its performance on both full-frame and region-level experiments, and, moreover, they claimed that the Multimodal RNN outperforms the retrieval baselines in both of them.

Feng et al. [36] proposed a robust kernel metric learning (RKML) algorithm based on the regression technique which can be directly utilized in image annotations. The RKML algorithm is also computationally more efficient due to the PSD feature is automatically ensured by regression algorithm.

Liu et al. [19] proposed a novel CNN-RNN image annotation model which utilizes a semantically regularized embedding layer as the interface between the CNN and RNN. However, they proposed semantic regularization that enables reliable fine-tuning of the CNN image encoder as well as the fast convergence of end-to-end CNN-RNN training. In practice, the semantic regularization generates the CNN-RNN interface semantically meaningful, and distributes the label prediction and correlation tasks between the CNN and RNN models, and importantly the deep supervision, i.e., it makes training the full model more stable and efficient.

Li et al. [24] introduced a global-local attention (GLA) method by combining local representation at object-level with global representation at image-level through attention mechanism. This method focuses on how to predict the salient objects more accurately with high recall while keeping context information at image-level. In the experimental results, they claimed that it achieved better performance on the Microsoft COCO benchmark compared with the previous approaches.

## III. Proposed Annotation Model

As depicted in Fig. 1, the overall structure of proposed model has three major sub-subsystems: the feature extractor, the tag generator, and the image-annotator. In the following, we will explain the basics, individual components, and their relationships in details.

## A. RNN and LSTM

Since the Recurrent Neural Network (RNN) [19], [22] and the Long Short Term Memory (LSTM) [23], [25], [32] are the basic components of the proposed annotation model, we describe briefly the RNN and the LSTM. A RNN networks at time $t$ reads a unit of the input sequence, which is a sequence of vectors such as $K = (k_1. k_2. ….k_T)$ and gets the previous state, $h_{t-1}$.

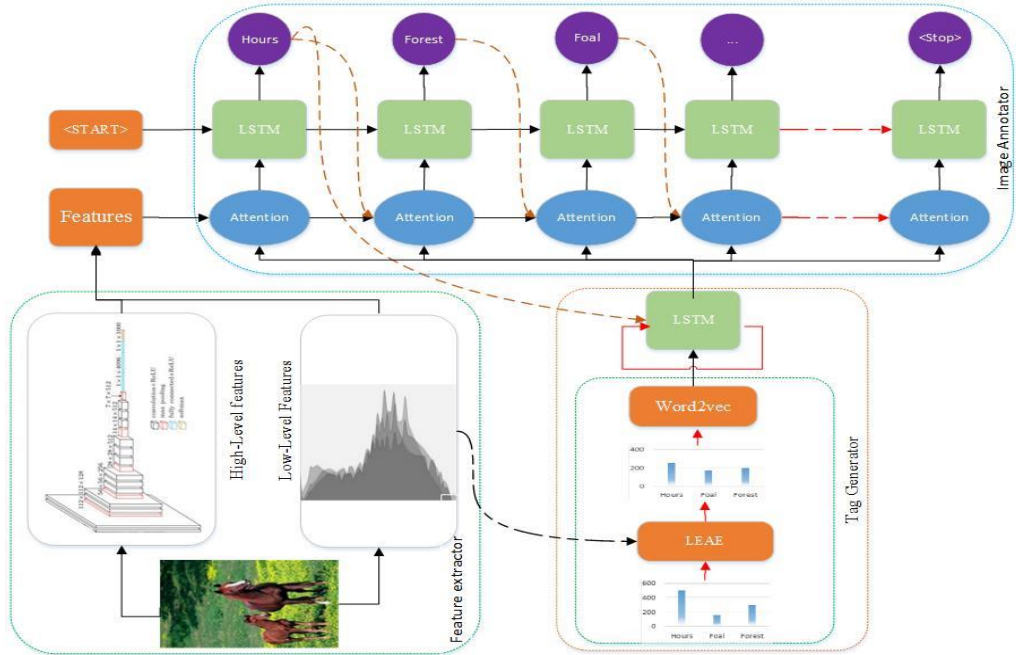

Fig. 1. The overall structure of proposed automatic image annotation model.

In addition, it generates the output and hidden layer at time $t$. The most common RNN approach can be calculated as follows:

$$h_t = f(k_t. h_{t-1}) \quad (1)$$

Where $f$ is a nonlinear function and $h_t$ is a hidden state at time $t$.

The LSTM is an advanced version of RNN with distinctive unit, which is able to manipulate the long-term dependencies. A LSTM unit consists of a cell state, and three gates (e.g., input, output, and forget gates). The input gate decides which values should be updated, while a sigmoid function does this operation. The forget gate decides what information should be removed from the cell state. Finally, the LSTM unit employs output gate by taking the same value with input and forget gates in order to obtain result based on the cell state. Fig. 2 shows the basic LSTM unit of the proposed model.

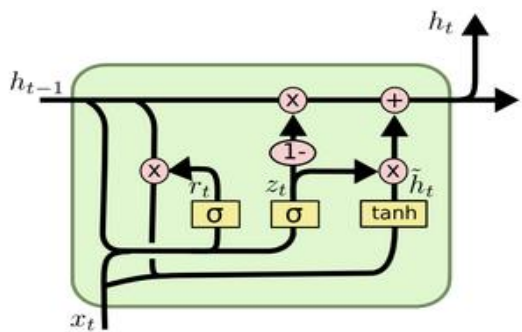

Fig. 2. Basic LSTM unit [48].

$$z_t = \sigma(w_z \times [h_{t-1}. x_t])$$
$$r_t = \sigma(w_r \times [h_{t-1}. x_t])$$
$$\widetilde{h_t} = \tanh(w \times [r_t \times h_{t-1}. x_t]) \quad (2)$$
$$h_t = (1 - z_t) \times h_{t-1} + z_t \times \widetilde{h_t}$$

## B. Problem Formulation

Let $X = \{I_1. I_2 …. I_N\}$. $I_i \in R^d$ denote $N$ training images that each. $\nabla = \{K_1. K_2. … K_M\}$ defines the dictionary of $M$ possible annotation keywords. if the $I_i$ annotated by the $Kj$, $\varphi(i.j) = 1$ Otherwise $\varphi(i.j) = 0$, and the $ith$ image is annotated by $\varphi_i = \{\varphi(i.1). \varphi(i.2). ….\varphi(i.M)\}$. The goal of image annotation is to select the appropriate set of tags for a given image. Image is represented by a d-dimensional features vector.

The upper LSTM of the proposed model used to train the images and predict the next appropriate annotation tag using a given attention vector. That $A_t$ is an attention output that can be calculated as follows:

$$A_t = g(A_{t-1}. TG. k'_{t-1})$$
$$A_1 = g(Attention\ weight. TG) \quad (3)$$

The $g$ is a nonlinear function and attention weight which obtains the high and low level features.

$$TG_t = \sigma(k'_1. TG_t. k_{t-1}) \quad (4)$$

$TG$ is the set of words that is generated by the tag generator sub-system. The $\acute{k}_1$ is the first annotation tag, which is generated by supervised training. In other words, it inputs to the tag generator and lead to production of words, which is closed to $\acute{k}_1$ and thus other sections of the proposed model select the appropriate tag.

*C. Feature Extractor*

In this section, we describe the feature extractor (low and high-level features) for image annotation in details. Low-level features and high-level features are essential for pattern recognition and associative learning (i.e., a proper combination of these features give a very important information from the details of the image, where the high-level features usually contain the context information about objects, and low-level features always contain context and background information) [26]. Therefore, we utilize both features for appropriate representation of the images. The detail structures of both features are described as follows.

*1) Low-level features*

Low-level image's features refer to a set of combination of elements or independent objects in an image [9]. These features are attributes that describes more specific individual components (e.g., color, shape, texture and background) of an image by focusing on the details of rudimentary micro information of images [15]. In this paper, we inspired the low-level features from the previous works in order to express the color and texture information of the image [37], [38].

The low-level features extraction method based on Distribution of Color Ton called "DCTon" works based on the structures of Texton methods [37], [39], [40]. Regular textures are the scanning result of images by the Texton component [39]. While the DCTon can be considered as the extended version of the Texton [38]. In the approach, the color connectivity regions of an image are considered as the proper properties for extract features that contain the color and texture information simultaneously. If $R, G, and B$ are the color components, and the DCTon components are utilized to describe a pixel with components $(R_1. G_1. B_1)$ (i.e., it appears a specified spatial relationship by considering distances and orientations of the corresponding pixel with components $(R_2. G_2. B_2)$ for each pixel of the image) [38]. After extracting the DCTon components, the values of pixels are set to their average in order to make the DCTon image. After creating the DCTon image, the Co-occurrence matrixes are extracted from this image. The Co-occurrence matrix of this quantized component is established and its contrast, correlation, energy, and homogeneity features are extracted by a vector which can be defined as follows:

$$LL_{DCTon} = [Contrast. Correlation. Energy. Homogeneity] \quad (5)$$

Another low-level features extraction method operates based on the Dual Tree-CWT and SVD (Support Vector Machine), and conceptual segmentation, which was introduced in [38]. These features extracted from n-levels decomposition of 2-D, DT CWT of a W×H input image. Each scale has a set of sub bands with size $\frac{W}{2^n}.\frac{H}{2^n}$ of complex coefficients, which can be denoted by (6):

$$M^i_{c.\,lowpass} = \begin{bmatrix} ll^i_{1.1} & \cdots & ll^i_{1.\frac{H}{2^n}} \\ \vdots & \ddots & \vdots \\ ll^i_{\frac{W}{2^n}.1} & \cdots & ll^i_{\frac{W}{2^n}.\frac{H}{2^n}} \end{bmatrix}.c$$
$$= \{real. imaginery\}. i$$
$$= 1.2. \quad (6)$$

$$M^j_{c.highpass}$$
$$= \begin{bmatrix} hh^i_{1.1} & \cdots & hh^i_{1.\frac{H}{2^n}} \\ \vdots & \ddots & \vdots \\ hh^i_{\frac{W}{2^n}.1} & \cdots & hh^i_{\frac{W}{2^n}.\frac{H}{2^n}} \end{bmatrix}.c$$
$$= \{real. imaginery\}. j = 1.2. \ldots .6.$$

For the conceptual segmentation of images, we divided the images into 5 regions ($A_1 to A_5$) same as the [38]. Moreover, we extracted the Real and the imaginary sub-band coefficients of four levels DT-CWT [14] decomposition from each image segment and calculated the SVD [18] of each of the derived matrixes using the vectors of eigenvalues such that the features of an image can be defined as (7).

$$LL_{tf} = \begin{bmatrix} A_1 & = [\sigma_1. \ldots . \sigma_{Rank(M_1)}. \sigma_1. \ldots . \sigma_{Rank(M_2)}. \cdots . \sigma_1. \ldots . \sigma_{Rank(M_{16})}] \\ \vdots & \vdots \\ A_5 & = [\sigma_1. \ldots . \sigma_{Rank(M_1)}. \sigma_1. \ldots . \sigma_{Rank(M_2)}. \cdots . \sigma_1. \ldots . \sigma_{Rank(M_{16})}] \end{bmatrix} \quad (7)$$

Where the $\sigma_1, \ldots, \sigma_i$ are the diagonal eigenvalues.

Therefore, the final low-level features of the image in the proposed model are a fusion of the DCTon based features and DT-CWT features, which can be denoted as follows:

$$LL = [LL_{DCTon}. LL_{tf}{}^T] \quad (8)$$

*2) High-level Features*

High-level features or thematic features are important attributes for image representation. These properties represent the image with a global perspective and refer to the definition of the image or concept of the image [20], [24]. Details of these features can imitate the human perceptual system very well. During the last decade, several approaches have been proposed to improve the high-level feature extraction in pattern recognition area. Convolutional Neural Network (CNN, or ConvNet) has recently utilized as a powerful class of models for feature extraction purpose such as VGG16, VGG19 and ResNet50 [16], [21]. In this study. We employ the VGG16, and the ResNet50 in order to extract the high-level features. If the ResNet models are at properly tuned. The feature extractor provides better results than other architectures. For example, the deeper ResNet with '34' layers has a smaller training error compared with the 18 layer. The ResNet model should utilize a direct path for propagating information through the residual block in the network as a result. It allows the information to propagate from one block to any other ones during both forward and backward passes. Due to this reason, it causes to reduce the complexity of the training process.

## D. Attention Mechanism

During the process of image annotation, some important features of attention are very essential (e.g., some features are important and some others are not). In practice, each feature has a special weight for generating the image representation. Herein, the way of fusing the low-level features and high-level features are important for images annotation. Due to this reason, we proposed an attention mechanism to integrate the high and low-level features so that it can selectively focus on some important objects of the image at different time. Moreover, it brings up the appropriate keywords at the same time using (9).

$$Attention\ weight = \sum_i^{\#(LL)} \gamma_i^{(t)} LL_i + \sum_j^{\#(HL)} \xi_j^{(t)} HL_j \quad (9)$$

Where $\gamma_i^{(t)}$ and $\xi_j^{(t)}$ denotes the attention weights of low and high-level features at time $(t)$, that

$$\sum_i^{\#(LL)} \gamma_i^{(t)} + \sum_j^{\#(HL)} \xi_j^{(t)} = 1 \quad (10)$$

$\gamma_i^{(t)}$ and $\xi_j^{(t)}$ can be calculated by the Softmax function [28],

$$\gamma_i^{(t)} = \tau \times \frac{\exp(\zeta_i^t)}{\sum_i^{\#(LL)} \exp(\zeta_i^t)}$$
$$\xi_j^{(t)} = (\frac{1}{2} - \tau) \times \frac{\exp(\varrho_j^t)}{\sum_j^{\#(HL)} \exp(\varrho_j^t)} \quad (11)$$
$$0 \le \tau \le 1$$

The attention weight at time $(t)$ has a direct relation to the previous information and the other features. Therefore, these relations can be modeled in a nonlinearity function. $\zeta_i^t$ and $\varrho_j^t$ can be defined as follows:

$$\zeta_i^t = \Gamma^T \sigma(W_h h^{t-1} + W_l \times LL_i + b)$$
$$\varrho_j^t = \Gamma^T \sigma(W_h h^{t-1} + W_l \times HL_i + b) \quad (12)$$

That $h^{t-1}$ is the previous hidden state.

Since $\zeta_i^t$ and $\varrho_j^t$ are not no rmalized, for normalizing these weights, we used the Softmax function and calculated the final weights. In addition, due to the numbers and dimensions of the low and high level features are different, there are various effects during annotation process. Therefore, we considered the $\tau$ coefficient for learning the proportional in order to define importance degree of the high and low features. Others parameters $\Gamma$. $W_l$ and $W_h$ are coefficients that can be learned by our attention model.

## E. Tags Generator

During the process of automatic image annotation, a sub-system that produces the cluster of same family words for image annotation is very important. For this purpose, as shown in Fig. 1, we proposed a tag generator which consists of two phases including, word embedding, and data equilibrium. The word embedding describes the keywords (as a text modal) by appropriate vectors and data equilibrium in order to apply to the imbalanced keywords and generates different weights for different frequency keywords. In addition, it balances the keyword dictionary. In the following, these two phases are described.

*1) Balanced/skewed distribution keywords*

In the image-annotation datasets, the keywords are extremely diverse with the skewed distribution, and the number of different keywords used for annotation of images is imbalanced. For example, the ESP game dataset [41], has over 20,000 images and 268 keywords. A high-frequency tag generator has been used for over 500 images, while a low-frequency tag is used for less than 50 images [13]. This problem extremely affects the performance of image annotation, and the low-frequency keywords have less effect during the annotation. As a result, the existing techniques have low percent accuracies, and conversely. Due to this, we utilize the imbalanced learning for generating tags with respect to variance distribution. In practice, it can increase the training intensity of low-frequency keywords for image samples in order to enhance the generalization performance of the whole model. For addressing this problem, we introduce an ANN based Auto-encoder method, which is used for unsupervised learning [2], [3]. The aim of an auto-encoder is to learn a representation for a set of data, typically for the purpose of dimensionality reduction. An auto-encoder always consists of two phases: the encoder $e_\theta$ and the decoder $d_{\theta'}$. The encoder transforms an input vector $k$ in to the hidden layer $h$. In addition, the decoder maps the $h$ back in order to reconstruct the input vector $k'$(e.g., reconstructed vector is optimized by the cost function).

$$e_\theta(k) = \sigma(w \times k + b) \ . \ \theta = \{w.b\} \quad (13)$$

Where W is weighted the matrix, $b$ is a bias vector, and $\sigma$ is a nonlinear activation function. Moreover, the decoder has the following relation.

$$d_{\theta'}(h)$$
$$= \begin{cases} \sigma(w' \times h + b') & when\ input\ vector\ \in [0\ 1] \\ w' \times h + b' & when\ input\ vector\ \in \mathbb{R} \end{cases} . \theta'$$
$$= \{w'.b'\} \quad (14)$$

Where $w' = w^T$ and $b' = b^T$

If the input vector is $k$ and the approximation of input vector is $k'$, the auto-encoder minimizes the loss function by the $|k - k'| < \mathcal{E}$. With respect to the mentioned relation, $k' = d_{\theta'}(e_\theta(k))$; and the auto-encoder model can be learned by the following optimization:

$$\theta^*.\theta'^* = \arg\ min\ \frac{1}{M}\sum_{i=1}^{M}|k_i - d_{\theta'}(e_\theta(k_i))| \quad (15)$$

Herein, the $M$ is the number of samples.

For the proposed balanced model, we consider $L$ as hidden layer. Let $h_l$ be the output vector of layer $l$, the feed-forward operation for $L$ layer of auto encoder can be described as follows:

$$h_{l+1} = \sigma(W^{l+1}h^l + b^{l+1}) \ . \ l\epsilon\{0. \ldots. L-1\} \quad (16)$$

The $h_0$ and $h_L$ are the input and output vectors using the following backpropagation algorithm.

$$\theta^* = \arg\min \sum_{i=1}^{M} |F_\theta(I_i).K_i| \quad (17)$$

Where $F_\theta(I)$ is a composition of activation function $\sigma$ from $\theta_1$ to $\theta_L$ and $\theta_l = \{W_l.b_l\}$ are the model parameters.

The low-level image features are the input to the model and the image tags are the supervision information. The backpropagation algorithm is used to create the relation between the features and tags. But the imbalanced distribution of the image keywords generates a model such that provides a skew degree of accuracies. To enhance the annotation performance, we propose a balanced Log-Entropy-Auto-Encoder (LEAE) that can enhance the training process for low frequency tags. We utilized the [8], [13], [42] for giving the appropriate coefficients to the tags using the concept of log-entropy. Assuming that there are $N$ images and $M$ different tags in the training dataset. If we increase the coefficients of low-frequency keywords, then it provides a balance training. For this purpose, we construct a coefficient matrix $W_{N \times M}$ for training all the images as follows:

$$W = \begin{bmatrix} w_{11} & \cdots & w_{1M} \\ \vdots & \vdots & \vdots \\ \vdots & \vdots & \vdots \\ w_{N1} & \cdots & w_{NM} \end{bmatrix}_{N \times M} \quad (18)$$

Each $w_{ij}$ can be calculated as follows:

$$w_{ij} = \varphi(i.j) \times \left(1 + \frac{1}{T_j \log N} \sum_{i=1}^{N} \varphi(i.j)[\log \varphi(i.j) - \log T_j]\right) \quad (19)$$

Where the $T_j$ is total number of occurrence of $jth$ tag in training images. The $(i.j) \epsilon \{0.1\}$, the $ith$ image is annotated by keyword $j$, $\varphi(i.j) = 1$ otherwise $\varphi(i.j) = 0$.

For example, assuming that, there are three images including:

$I_1.I_2$ and $I_3$ and three keywords in the training dataset, $K_1.K_2$ and $K_3$ are the annotation tags are assigned as depicted in Table I.

TABLE I. EXAMPLE OF IMAGE AND ANNOTATION TAGS

| Images | $I_1$ | $I_2$ | $I_3$ |
|---|---|---|---|
| Annotation tags | $K_1.K_2$ and $K_3$ | $K_2$ and $K_3$ | $K_3$ |

Note that the $K_1$ is the low-frequency tag.

The $w_{12} = 0.36$ and $w_{11} = 1$.

Therefore, we train the proposed model by the following optimization in order to balanced learning.

$$\theta^* = \arg\min \sum_{i=1}^{M} |F_\theta(I_i).W_i \times K_i| \quad (20)$$

Where $W_i$ denotes the $ith$ row of coefficient matrix $W$.

*2) Word-embedding descriptor*

In the task of AIA based on modals data such as image and text, there are many kinds of relations that are included of image-to-image, image-to-word, and word-to-word relations [5]. We address these mutual modal relations between image and word due to it is very important during the annotating process. In this sub section, we will describe the relation of word-to-word.

If we employ a proper description of the words, then the quality of the tag generator, the consequently, and the quality of the AIA system will be increased. In the both situations, if the words semantically are similar, and described with similar vectors; or the words semantically are not similar described by vectors with proper semantic distance; the tag generator can suggest the best words to other sub-system of model.

In general, the word embedding approaches are used in the natural language processing in order to transform the bag-of-words representation to a continuous space representation [28]. There are some advantages to this continuous space since the dimensionality is largely reduced and the words closer in meaning are close in this new continuous space. There have been introduced some applications of the word embedding based on neural networks including the word2vec [26], Dictionary of Affect in Language (DAL) [44], SentiWordNet [43], Glove [26] and Wikitionary [45].

We have used the word2vec [26], which offer two possible ways to generate the word embedding continuous bag-of-words- and skip-gram (CBOW). After training the word embedding, an n-dimensional vector is available for each word in the dictionary. In order to training the Word2Vec, we utilized a Large Textual Corpora such as all Wikipedia articles in a certain language. Moreover, we applied the thematic textual collections related to semantic concept collections, and annotation keywords during the Word2vec training.

## IV. EXPERIMENTAL RESULTS AND ANALYSIS

In this section, we implemented the proposed model on two different datasets in order to evaluate the performance of the proposed model. First, we introduce a detailed description of the datasets and evaluation metrics. Then, we compare the experimental results with the state-of-the-art techniques.

*A. Datasets*

The Corel-5k dataset is one of the famous benchmarks that have been used for evaluating the AIA models so far. Herein, we used this benchmark in order to analyze and compare the proposed model with other models. The Corel dataset consists of 5000 images from 50 Corel Stock Photo CDs, and each CD includes 100 images on the same topic, an image is manually labeled at least one-annotation word and maximum of five-annotation words [3]. All of distinct tags in dictionary are 260 keywords. The training set consists of 3500 images, validation set includes 750 images and test set contains 750 images.

The IAPRTC-12 dataset consists of 19,627 images of sports, actions, people, animals, cities, landscapes and many other aspects of contemporary life. All of distinct tags in dictionary are 291 keywords. The training set consists13739 images, the validation set contains 2944 images and the test set consists of 2944 images.

*B. Evaluation Metrics*

We have implemented the experiments in the Python 3 programming using the *TensorFlow*, *NumPy,* and *Keras* frameworks and run on the same *PC* with *Intel Centrino Core*

*i7 2670QM Duo 2.20 GHz* Processor, *8GB* of RAM, and Linux Ubuntu 14.1 operating system.

Since the common evaluation metrics can be easily found in recently proposed models [10]-[20], we utilized them in order to evaluate the performance of the proposed model. Moreover, we computed the evaluation metrics including the precision, and recall for each keyword separately (i.e., first, five relevant tags are extracted for each image and all images are annotated with these tags; Second, for each keyword $k_j$, average recall and precision rate are computed by the following equations). It is assumed that $K_j \epsilon \nabla$ and $\nabla$ is the dictionary of $M$ possible annotation keywords.

$$\overline{Recall} = \frac{\sum_{j=1}^{M} Precision(k_i)}{M \sum_{j=1}^{M} R(k_i)} \quad (21)$$

$$\overline{Precision} = \frac{\sum_{j=1}^{M} precision^j}{M \sum_{j=1}^{M} Prediction(k_i)} \quad (22)$$

That $Precision(k_i)$ is total number of the correctly predicted keyword $k_i$, $R(k_i)$ is the relevant annotated counts of the keyword $k_j$ and $Prediction(k_i)$ is the predicted counts of the keyword $k_j$. The "$F - measure$" is a measure of the test accuracy, which considers both the $\overline{precision}$ and the $\overline{recall}$ of the test in order to compute the score. This measure can be calculated as follows:

$$F - measure = \frac{2 \times \overline{precision} \times \overline{recall}}{\overline{recall} + \overline{precision}} \quad (23)$$

*C. Comparison Results*

In order to show the superiority of the proposed image annotation model, we compare it with several state-of-the-art models, including L-RBSAE [8], Multi-AE [13], G-LSTM [17], 2PKNN [46], JEC [47] and Soft/Hard Attention [20] by implementing on Corel-5 K and IAPRTC-12 datasets. In practice, there are some differences between the mentioned models. The first difference is the feature extraction sub-system (i.e., the G-LSTM uses GoogLeNet, to extract the features; Multi-AE employs multi-view including the GIST [8], Hue [29] and SIFT [30] in order to extract the features; The JEC exploits the AlexNet to extract image features; The 2PKNN and Soft/Hard Attention emp loys the VGG16 as the same like our model to extract image features). The second difference is that the structure of image annotation (i.e., the Multi-AE utilizes the basic RNN as the decoder for the annotating process; the G-LSTM and Soft/Hard Attention utilizes the LSTM network for the annotating process).

The evaluated results of proposed model and other mentioned models are listed in the Table II. As we already pointed out, the performance analysis of the proposed model and each of the state-of-the-art models are evaluated based on the accuracy measure $\overline{precision}.\overline{recall}$ and $F - measure$. As depicted in Table II, the bold numbers indicate the performance of the proposed model and it is obvious that the proposed model noticeably outperforms most of the evaluated models, especially the Multi-AE, the JEC and the L-RBSAE.

Meanwhile, it is a little better than Soft/Hard Attention and G-LSTM.

TABLE II. PERFORMANCE COMPARISON OF PROPOSED MODEL VS THE EVALUATED MODELS

| Datasets | Model | $\overline{precision}$ | $\overline{recall}$ | $F - measure$ |
|---|---|---|---|---|
| Corel5k | Multi-AE[13] | 0.15 | 0.2 | 0.17 |
| | 2PKNN[46] | 0.22 | 0.66 | 0.33 |
| | JEC[47] | 0.18 | 0.47 | 0.26 |
| | L-RBSAE[8] | 0.23 | 0.24 | 0.23 |
| | G-LSTM [17] | 0.22 | 0.72 | 0.33 |
| | Soft/Hard attention[20] | 0.22 | 0.75 | 0.34 |
| | **Proposed Model** | **0.28** | **0.96** | **0.43** |
| IAPRTC-12 | Multi-AE[13] | 0.43 | 0.38 | 0.40 |
| | 2PKNN[46] | 0.30 | 0.38 | 0.34 |
| | JEC[47] | 0.29 | 0.19 | 0.23 |
| | L-RBSAE [8] | 0.28 | 0.63 | 0.38 |
| | G-LSTM [17] | 0.35 | 0.57 | 0.43 |
| | Soft/Hard Attention[20] | 0.40 | 0.48 | 0.43 |
| | **Proposed Model** | **0.54** | **0.37** | **0.44** |

Obviously, the evaluated results confirm that the proposed model provides almost the same performance results like the G-LSTM and Soft/Hard Attention. In addition, we can observe that the significant performance improvement compared to the Multi-AE or the L-RBSAE. The precision and the recall metrics of proposed model are also comparable with those of the recently proposed models, for the Corel-5 K and IAPRTC-12 datasets. Fig. 3 shows some sample images and annotations examples from the Corel dataset after implementing the proposed model. Even though, some of the predicted tags for these annotations do not match with the image, still they are very meaningful.

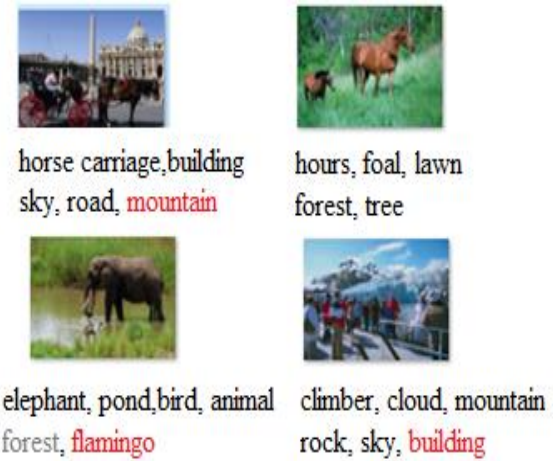

Fig. 3. Qualitative image annotation results obtained with our model.

## V. Conclusions and Future Work

In this paper, we have presented a novel AIA model based on the deep learning feature extraction method. In addition, we proposed a loge entropy solution in order to solve the problem of imbalanced data in image annotation. First of all, we implemented the proposed model on two popular datasets and, second, we evaluated the obtained results with respect to evaluation metrics. Finally, we compared the proposed model with several state-of-the-art models. The evaluated results confirm that the proposed model provides efficient performance and outperforms the evaluation metrics compared to the evaluated state-of-the-art models.

As for future work, we plan to design an efficient AIA model in order to improve the relevance score of features and utilize the combining generative and discriminant methods to improve the performance of the new AIA model.